\documentclass[12pt]{article}
\usepackage[english]{babel}
\usepackage[a4paper,top=2cm,bottom=2cm,left=2cm,right=2cm,marginparwidth=1.75cm]{geometry}
\usepackage{authblk}
\usepackage{graphicx}
\usepackage{newtxtext}
\usepackage{newtxmath}
\usepackage[square,sort,comma,numbers]{natbib}
\usepackage{hyperref}
\usepackage{wasysym}
\usepackage{float}
\usepackage{soul}

\hypersetup{
    colorlinks = true,
    urlcolor   = blue,
    citecolor  = black,
}

\newcommand{\RomanNumeralCaps}[1]
\linenumbers
\newcommand{\ora}{\overrightarrow}

\title{H$_2$ bubble motion reversals during water electrolysis
}

\author[1,2,3]{A. Bashkatov\footnote{E-mail: a.bashkatov@hzdr.de}}
\author[1]{A. Babich}
\author[1]{S. S. Hossain}
\author[1]{\\X. Yang}
\author[1]{G. Mutschke}
\author[1,2,3]{K. Eckert\footnote{E-mail: k.eckert@hzdr.de}}

\affil[1]{Institute of Fluid Dynamics, Helmholtz-Zentrum Dresden-Rossendorf, Bautzner Landstrasse 400, Dresden, 01328, Germany}
\affil[2]{Institute of Process Engineering and Environmental Technology, Technische Universit{\"a}t Dresden, Dresden, 01062, Germany}
\affil[3]{Hydrogen Lab, School of Engineering, Technische Universit{\"a}t Dresden, Dresden, 01062 Germany}

\date{}

\begin{document}

\maketitle

\vspace{-1cm} 
\begin{abstract}
The dynamics of hydrogen bubbles produced by water electrolysis in an acidic electrolyte 
are studied using electrochemical and optical methods. A defined cyclic modulation of the electric potential is applied at a microelectrode to produce pairs of interacting H$_2$ bubbles in a controlled manner. Three scenarios of interactions are identified and systematically studied.
The most prominent one consists in a sudden reversal in the motion of the first detached bubble, its return to the
electrode and finally its coalescence with the second bubble.
Attested by Toepler's schlieren technique,
an explanation of contactless motion reversal is provided by the competition between buoyancy and thermocapillary effects.
\end{abstract}

\section{Introduction}
\begin{figure}\centering
	\includegraphics[width=0.9\textwidth]{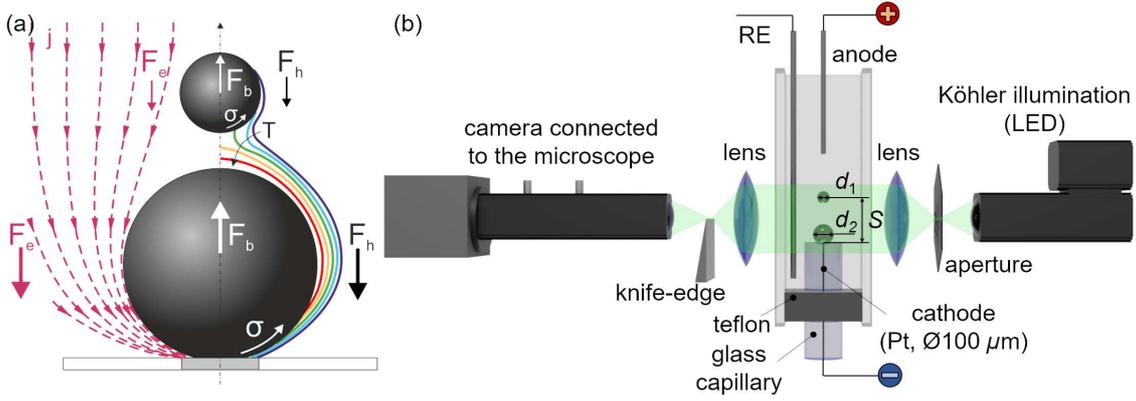}
	\caption{(a) Schematic of the pair of H$_2$ bubble produced by the current density j (dashed lines of red arrows) together with the forces acting on the bubbles (cf. Section 1). The 
	contour
	lines on the right represent the isotherms, 
	which decay from red to violet.
	(b) Scheme of the three-electrode electrochemical cell and the optical systems. $d_1 = 2R_1$ and $d_2 = 2R_2$ are the diameters of the first and second bubbles, and $S$ denotes the distance between the center of the first bubble and the electrode surface.
		\label{fig:scheme}}
\end{figure}

The growth and detachment of nano- and micrometer gas bubbles are omnipresent phenomena in nature and engineering, e.g. see the review by \cite{lohse2018bubble}. 
The growth of the gas bubbles in alkaline water electrolysis is a particularly interesting problem of high practical relevance.
Although alkaline water electrolyzers are
the most mature technology, they still suffer from low efficiency (\cite{smolinka2021electrochemical}) as a considerable part of the losses are caused by generated gas bubbles that block electrocatalytic sites and also raise the Ohmic cell resistance (\cite{angulo2020influence}). 
Thus, the rapid and efficient detachment of the bubbles from the electrodes is important; it is closely linked to a better understanding of 
the balance of forces acting on the bubble, the concept of which is comprehensively covered by \cite{thorncroft2001bubble}.
Recently, progress has been made in identifying and quantifying the forces of attraction on H$_2$ bubbles that counteract their buoyancy. Investigating the growth of H$_2$ bubbles under extreme cathodic potentials in acidic electrolytes, \cite{bashkatov2019oscillating}  and \cite{hossain2022} attributed the 
positional oscillations of H$_2$ bubbles prior to detachment to the action of two forces. First, the electric force $F_e$, given by
\begin{equation}
    \ora{F_e}=\int_\mathcal{S}\sigma E_z\,\mathrm{d}A
\end{equation}
where $E_z$ is the vertical ($z$) component of the external electric field, directed from the anode to the cathode, see Fig. \ref{fig:scheme}(a). $\mathcal{S}$ is the interface between the bubble and the electrolyte, and
$\sigma$ is the corresponding surface charge density, which is positive for gas bubbles in acidic electrolytes below the iso-electric point at pH $<$ 2...3  (\cite{brandon1985growth}).
The second important force is the hydrodynamic force $F_h$, given by (\cite{meulenbroek2021competing,hossain2022})
\begin{equation}
      \ora{F}_h=\ora{F}_M+\ora{F}_n=\int_\mathcal{S}\ora{\tau_h}\, \mathrm{d}A = -\int_\mathcal{S}\ora{\tau_M}\, \mathrm{d}A+\int_\mathcal{S}\ora{\tau_{h,n}}\, \mathrm{d}A
  \label{eqn:hydrodynamic-force}
\end{equation}
which is obtained by integration of the stress tensor, $\overrightarrow{\tau_h}=-p_h\ora{n}+\mu\frac{\partial \overrightarrow{u}}{\partial n}+\mu\nabla u_n$ over the bubble surface.
$p_h$ is the hydrodynamic pressure, $\mu$ is the dynamic viscosity of the electrolyte, $\ora{u}$ is the electrolyte velocity vector, $\ora{n}$ is the surface-normal unit vector, and $u_n=\ora{u}\cdot\ora{n}$.
$F_h$ originates in the fact that the surface tension ($\gamma$) of the gas-electrolyte interface depends on the 
temperature $T$ and/or species concentration $c$. Any gradient in $T$ or $c$ causes a gradient in  $\gamma$.
This gradient in $\gamma$ generates an imbalance in the shear stress that causes 
bubble surface elements, and the nearby electrolyte, to move from high to low $\gamma$ regions
(see \cite{lubetkin2003thermal,kassemi2000steady,guelcher1998thermocapillary,hossain2020thermocapillary}).
$F_M = -\int_\mathcal{S}\ora{\tau_M}\, \mathrm{d}A$ and $F_n = \int_\mathcal{S}\ora{\tau_{h,n}}\, \mathrm{d}A$ capture the contribution from the tangential stress that leads to the Marangoni convection, and the contribution from the normal stress (see \cite{hossain2022}).
Previous work by \cite{massing2019thermocapillary} provided evidence that the dominant source of the Marangoni convection observed at the bubble foot is thermocapillary rather than solutocapillary (see \cite{yang2018marangoni, meulenbroek2021competing, hossain2020thermocapillary}). 

Despite this progress, a number of unresolved phenomena remain, such as the bubble jump-off after the coalescence of two bubbles and its subsequent reattachment to the electrode (see \cite{westerheide1961isothermal, janssen1970effect, hashemi2019versatile}). There has been speculation as to whether electrostatic
attraction, Marangoni effects (\cite{lubetkin2002motion}) or coalescence, as recently proposed by \cite{hashemi2019versatile}, are behind the physics of the reattachment.

The focus of the present work is to achieve a better understanding
of such bubble interaction phenomena. As will be shown, this can
be achieved by more closely examining the phenomenon of thermocapillary migration. 
\cite{young1959motion} and \cite{hardy1979motion} were the first to show that an air bubble subjected to a vertical temperature gradient can move downward against the direction of buoyancy if
the liquid is heated from the bottom.
Experiments performed on a NASA Space Shuttle in orbit (\cite{balasubramaniam1996thermocapillary}) reported a migration velocity of $\sim 0.3$ mm/s in a 1 K/mm temperature gradient for air bubbles of $\sim$7...10 mm in diameter, in good agreement with the theory developed by \cite{young1959motion}.
A further type of bubble motion against buoyancy is the periodical bouncing of a plasmonic bubble in a binary liquid (\cite{zeng2021periodic}) as a result of competition between soluto- and thermocapillary effects. 
By interrupting continuous laser irradiation during the bubble growth on photocatalytic surfaces
\cite{cao2020visualization, cao2022regulation}
recently succeeded in forcing bubbles to take on a bouncing motion, during which they detach and reattach at the photocatalyst's surfaces. The re-attachment has been attributed to a thermal Marangoni effect.

In this work we add to previous studies by \cite{bashkatov2019oscillating} 
by including modulation cycles of the cathodic potential.
This modulation enables consecutive pairs of hydrogen bubbles of a well-defined size to be produced, which further allows the forces $F_e$ and $F_h$ to be varied. In this way we are able to systematically study the phenomenon of bubbles returning towards the electrode.
By using Toepler's schlieren technique alongside shadowgraphy and PTV, we were able to identify the origin of the H$_2$ bubble motion reversal as a kind of self-organized thermocapillary migration provoked by the interaction of two H$_2$ bubbles.  

\section{Experimental setup and procedure}
Consecutive
single H$_2$ bubbles were generated during water electrolysis in 0.5 M H$_2$SO$_4$ at a $\diameter$100 $\mu$m Pt microelectrode acting as a cathode, see Fig. \ref{fig:scheme}(b). Two Pt wires served as the anode and pseudo reference electrode, respectively. 
The cathodic potential, $E$, is modulated 
over time as shown in Figure \ref{fig:phen1}(a) to study the bubble-bubble interaction.
The modulation cycle of $E$ consists of three phases. In phase 1, the potential $E_1$ is applied for a short time $t_1$ to produce the first bubble,
which grows up to a radius $R_1$.
In the following phase "0" with a duration of $t_0$, the potential is switched off, i.e., $E = 0$.  
As this leads to the decay of the retarding forces
$F_e$ and $F_h$, it allows the first bubble to detach from the electrode 
and to rise over the distance $S$.
In the subsequent phase 2, the cathodic potential is switched on again at a larger value $E_2$  for a time period $t_2$. In this phase, a second bubble quickly grows, thereby possibly interacting with the first bubble if it is still close enough. Finally, the potential is again set to $E=0$ over a longer timespan $t_w$ to allow the bubbles that are produced to detach and the resulting electrolyte flow to decay. After this, 
the next cycle is initiated. A large number of such cycles, e.g. 105 
in Fig.~\ref{fig:Fig3}(d) and 133 in Fig.~\ref{fig:Fig3}(e), 
have been studied 
to ensure that the statistics
for the results reported are robust.

The experiments were performed at $E_1 = -2...-6$ V and $E_2 = -8$ V applied for $t_1$ = 1...5 ms and $t_2$ = 40...200 ms, respectively, while  interruption times $t_0$ of 120...200 ms
were applied for the detachment and rise of the first bubble. The waiting time between subsequent cycles was chosen as $t_w$ = 500 ms. 
This modulation scheme allows (first) bubbles of a very defined radius of e.g. $R_1=(66\pm 1)\, \mu$m to be produced, as in Fig.~\ref{fig:phen1}. These travel over a distance $S_{E2}$ before the second bubble is produced.

A high-speed shadowgraphy system (resolution: 1000 pix/mm, frame rate: 5 kHz) was used to visualize the bubble dynamics, as already described in \cite{bashkatov2021dynamics}. Monodisperse polystyrene particles ($\diameter$5 $\mu$m, $\rho_{ps} = 1.05$ g/cm$^3$) were seeded into the electrolyte to study the electrolyte flow by means of particle tracking velocimetry (PTV). For that purpose, the particle's path, acquired over 16 images per time instant (corresponding to 3.2 ms at 5 kHz) was additionally averaged over the 105 cycles.
The optics are further complemented by a Toepler's schlieren system (\cite{settles2001schlieren}) consisting of an aperture stop, two lenses  (focal length $f$ = 100 mm) and a horizontally installed knife edge to map the vertical refractive index gradients $\partial n/\partial z= dn/dT \cdot \Delta T/\Delta z + dn/dc \cdot \Delta c/\Delta z $ accompanying the evolution of electrogenerated H$_2$ bubbles. 
To enhance the contrast and signal-to-noise ratio, each schlieren image is divided pixel by pixel by a plain schlieren image without bubbles (\cite{huang2007modified}). 
As $dn/dT \cdot \Delta T/\Delta z\sim 10^{-3}$ (water, $\Delta T \sim 10$ K) (\cite{haynes2016crc}) while $dn/dc \cdot \Delta c/\Delta z\sim 0.5 \times 10^{-5}$ (air-saturated vs degassed water)  (\cite{harvey2005effect}), the
refractive index gradient maps the temperature rather than the concentration gradient. 

\section{Results}
\subsection{Spontaneous H$_2$ bubble motion reversal  opposite to buoyancy – Scenario I}
\label{sec:phendescr}
\begin{figure}\centering
	\includegraphics[width=0.9\textwidth]{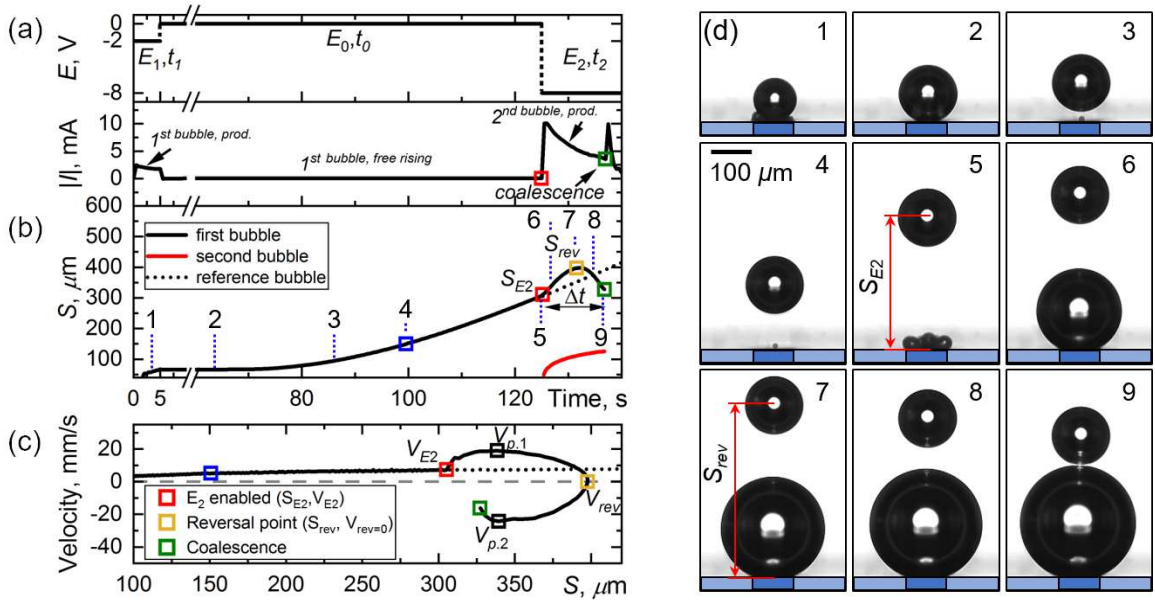}
	\caption{(a) 
 The cathodic potential $E$ modulated over time and the modulus of the resulting electric current $|I|$. (b) Distance $S$ between the center of the first bubble and the electrode over time. (c) Velocity $V$ of the first bubble versus distance $S$. The dashed line marks $V=0$, the dotted one corresponds to 
a continuously rising bubble for unmodulated, constant $E$.
 (d) Snapshots of the bubbles' behavior
 at time instants labeled in subfigure (b).
		\label{fig:phen1}}
\end{figure}

Figure \ref{fig:phen1} describes the basic phenomenon studied in terms of $E$ and $I$ (a), the distance to the electrode $S$ (b), the velocity of the first bubble (c) and snapshots of the H$_2$ bubble(s) (d) during the modulation cycle of the potential $E$.
In phase 1 (potential $E_1=-2$ V), the first bubble is produced. It already has a radius of around $R = 54\:\mu$m after 2.6 ms (snapshot 1), and reaches a final size of around $R = 66 \pm 1\:\mu$m after $t_1=5$ ms. 
In phase "0", where the potential is set to $E=0$, the hydrogen evolution reaction stops. The bubble resides at the electrode for a short time before detaching, depicted by snapshot 2 at 65 ms. After detaching, the bubble performs a free rise (snapshots 3–4). 
When phase 2 begins, the potential is switched to $E_2=-8$ V (snapshot 5) and the second bubble is produced. From now on, a completely unexpected process sets in. After the initial acceleration, the first bubble starts to decelerate (snapshots 5–6). At a distance $S_{rev}$ (snapshot 7), it finally reverses its direction of motion. Without any external influence, the bubble henceforth moves {\it against buoyancy} towards the second bubble, and coalesces with it (snapshots 8–9).

Figure \ref{fig:phen1}(b) analyses this phenomenon in terms of the distance $S$ between the center of the first bubble and the electrode surface. As long as the bubble is attached to the electrode, $S=R_1$. After it detaches and during the short acceleration phase of the first bubble, $S$ increases nearly
linearly with time between snapshots 4 and 5. When the second bubble appears (see red curve), upon switching to $E_2$ at $S=S_{E2}$, $S$ increases at a higher rate until a maximum $S_{rev}$ is attained. Thereafter, the bubble motion occurs in the reverse direction, and $S$ decreases until the two bubbles coalesce (9).
 

The velocity $V=dS/dt$ of the first bubble during these stages is plotted in Figure \ref{fig:phen1}(c).  Until the point of motion reversal at $S_{rev}$, the velocity of the first bubble is positive, and negative afterwards. 
When $E_2$ is switched on, the quickly growing second bubble displaces the electrolyte. This accelerates the first bubble upwards, and its velocity increases from $V_{E2}$ to a maximum of $V_{p.1}$ = 18.9 mm/s (black square, attained at snapshot 6). After that point, the interaction between the two bubbles' first forces the velocity of the first bubble to decrease to $V=0$ at $S_{rev}$. Afterwards, the first bubble is accelerated against buoyancy towards the second bubble. Hence, a second peak in the velocity, $V_{p.2}$ = -24.3 mm/s, the magnitude of which is larger than $V_{p.1}$, is attained shortly before coalescence with the second bubble.

\subsection{Full parameter space: Scenarios I–III}
\begin{figure}\centering
	\includegraphics[width=0.9\textwidth]{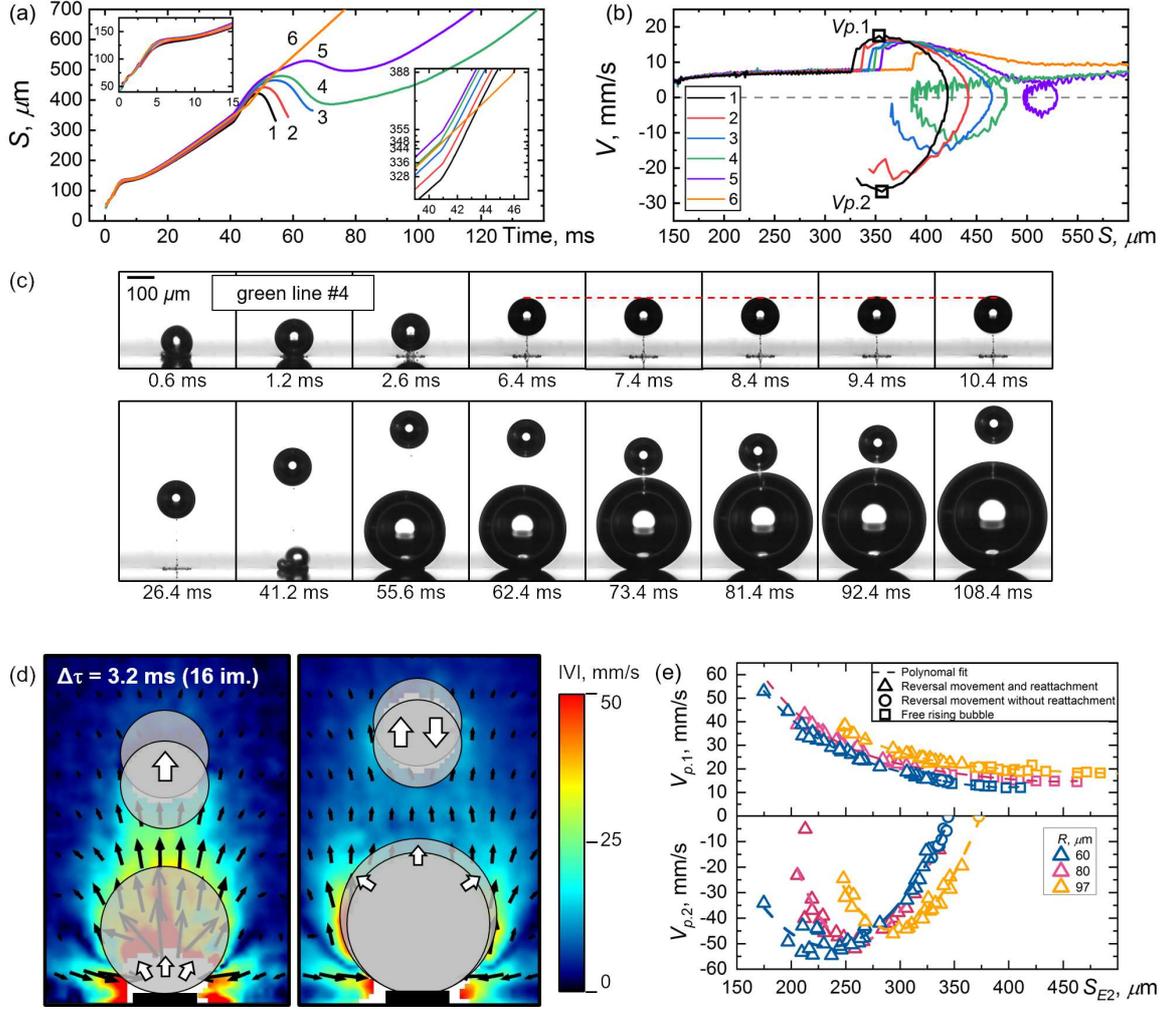}
	\caption{Overview of the three scenarios in terms of the position $S$ of the first bubble over time (a) and its velocity $V(S)$ (b). 
	Lines 1 to 3 are scenario I, lines 4 and 5 are scenario II, and line 6 is scenario III. The insets in (a) show $S(t)$ shortly after the bubble nucleation (left) and close to the time $E_2$ is switched on (right). (c) Bubble snapshots belonging to line 4 (scenario II).
	 	(d) Velocity magnitude contour and vectors in the electrolyte during the growth of the second bubble (left) and at the point of reversal of the first bubble at $S_{rev}$. 
	(e) 
	Velocity peak values $V_{p.1}$ and $V_{p.2}$ versus distance $S_{E2}$ for three different bubble radii (different colors). Triangles, circles and squares relate to scenarios I, II and III, respectively.
		\label{fig:Fig3}}
\end{figure}

The occurrence of scenario I, described above, depends crucially on the first bubble's distance $S$ at which $E_2$ is switched on.
Figure \ref{fig:Fig3} supplements Figure \ref{fig:phen1} by plotting the three possible scenarios in terms of the position of the first bubble over time $S(t)$ (a) and its velocity $V(S)$ (b). Each line, labeled from 1 to 6, represents the trajectory of the first bubble for six different cycles. The experiments were performed at $E_1 = -6$ V, applied for $t_1=5$ ms, and $E_2 = -8$ V, applied for $t_2=40$ ms (1$^{st}$ to 5$^{th}$ bubble) and 45 ms (6$^{th}$ bubble).
The radius of the first bubble is determined by $E_1$ and $t_1$. As both values were identical for all 6 cycles, the bubbles have nearly the same radius $R=66 \pm 1$ $\mu$m. However, the tiny variations in $R$ already cause slight deviations in the detachment process, see left inset in (a), and Table 1. 
As such, the distances $S_{E2}$ at the onset of $E_2$ vary, 
as seen in the right inset in subfigure (a), where the different bubble paths are also labeled by numbers.

The scatter in $S_{E2}$, although being smaller than 30 $\mu$m in size, 
plays a crucial role in deciding which scenario the bubble follows later on.
For small distances $S_{E2}< S_{E2.crit}$, the first bubble follows scenario I (lines 1...3) as described in Section \ref{sec:phendescr}.
For larger distances $S_{E2.crit} \leq S_{E2} <  S_{E2.6}$, a different scenario II is found (lines 4 and 5), which is visualized in subfigure (c). 
Here, a reversal of the motion of the first bubble at a distance $S_{rev}$ 
is also observed. However, the values of $S_{rev}$ are shifted to somewhat higher values compared to scenario I. 
The subsequent downward acceleration of the first bubble towards the second one
is comparably weaker and not sufficient to force the two to coalesce. Hence, upon approaching the second bubble, the first bubble 
is repelled and continues its rise.
Line 5 corresponds to a case of even slower downward motion.

The transition between scenario I (line 3) and scenario II (line 4) occurs at approximately $S_{E2.crit} \approx (S_{E2.3}+S_{E2.4})/2=346$ $\mu$m during which the
difference  $\Delta S_{E2}=S_{E2.4}-S_{E2.3}$
amounts to 4 $\mu$m only. 
At larger distances $S_{E2} \ge S_{E2.6} \sim 388\:\mu$m, the motion of the first bubble is only affected by displaced electrolyte during the fast growth of the second bubble.
However, deceleration does not occur that leads to a reversal of the direction of motion.  This corresponds to scenario III, represented by line 6. 

Two snapshots of the velocity field, obtained by PTV (cf. Section 2), are documented in Figure \ref{fig:Fig3}(d). Images of the bubbles, drawn to scale, are schematically superimposed. White arrows indicate the direction of the expansion and the motion of the bubbles, respectively. The velocity by which the growing second bubble
displaces the surrounding electrolyte scales with the bubble's growth rate and decays with distance $S$.
Thus, (first) bubbles at a higher $S_{E2}$ experience a smaller advection by the displacement flow.  
As the bubble's growth rate decreases with time, the velocity of the displacement flow also decreases,  see the differences between left and right images in Figure 3(d).
Although this flow is still directed upward in the right image, which shows the bubble at $S_{rev}$, the bubble starts to reverse the direction of its movement, and is accelerated towards the electrode.
We further note that the high magnitude of the velocity visible at the foot of the second bubble (Figure \ref{fig:Fig3}(d), right image) is caused by the temperature gradient along the bubble surface arising from Joule heating due to the high current density at the rim of the microelectrode (\cite{yang2018marangoni}). 
The resulting Marangoni convection 
is further enhanced by the bubble expansion.

The characteristic velocity maxima, $V_{p.1}$ and $V_{p.2}$, attained by the first bubbles in Figure \ref{fig:Fig3}(a) are summarized in Table \ref{tab:1} for the three different scenarios represented by lines 1 to 6. The velocity $V_{E2}$ of the first bubbles
at the onset of potential $E_2$ is also included. It is 
interesting to compare
all three quantities to the terminal velocity $V_t$ of a freely rising bubble (\cite{clift2005bubbles}). For the present case ($R = 66\:\mu m$, Reynolds number Re $\approx$ 5, Eötvös number Eo $\ll$ 1), $V_t$ can be estimated as amounting to
 $   V_{t} = \frac{2R^2\Delta\rho g}{9 \mu} \sim 8.8$ mm/s 
where $\Delta\rho$ and $\mu$ denote the density difference and the dynamic viscosity. 
Inspecting Table \ref{tab:1}, we note that all $V_{E2}$ values are $7...16\%$ smaller than $V_t$. The main reason is that the bubbles have not yet finished the initial acceleration phase at the comparatively short distances of $S<400 \, \mu$m reached
before $E_2$ is switched on. For that reason, $V_{E2}$ rather than $V_t$ is used as a reference velocity to non-dimensionalize $V_{p,1}$ (cf. Table \ref{tab:1}). 
On examining $V_{p.1}/V_{E2}$, we see that $V_{p.1}$ exceeds the velocity of the freely rising bubble by a factor of 1.7 to 2.4. The reason is that the bubble's velocity with respect to a non-moving frame results from the superposition of free rise and advection by the "bow wave" of the displaced electrolyte.  
The high magnitudes of $|V_{p.2}|$ attained by the bubble during its reverse motion against buoyancy are even more noticeable. They may exceed $V_{p.1}$ by up to a factor of 1.5 in scenario I. 
\begin{table*}
	\centering
	\small
	\begin{tabular}{cccccccc}
		Scenario / & $S_{E2}$ & $\Delta t$ &$V_{E2}$ &$V_{p.1}$ &$\mid V_{p.2} \mid$ &$V_{p.1}/V_{E2}$ &$|V_{p.2}|/V_{p.1}$ \\
		Line, \verb|#|&  mm & ms &  mm/s & mm/s & mm/s & &\\
		\hline
		I / 1 &328 &13.2 &7.4 &17.5 &26.7 &2.37 &1.53 \\
		I / 2 &336 &17.2 &7.8 &16.3 &23.3 &2.09 &1.43 \\
		I / 3 &344 &25& 8.0 &16.0 &15.9 &2.00 &0.99 \\
		II / 4 &348 &--- &7.9 &16.0 &12.7 &2.03 &0.79 \\
		II / 5 &355 &--- &8.0 &15.8 &5.7 &1.98 &0.36 \\
		III / 6 &388 &--- &8.2 &13.9 &--- &1.69 &--- \\
	\end{tabular}
	\caption{Summary of the characteristic values from Figure \ref{fig:Fig3} for the three different scenarios.
	$V_{E2}$ refers to the free rise velocity of the first bubble at distance $S_{E2}$. $\Delta t$ is the time interval
	between instants of time 5 and 9 as marked in Fig. \ref{fig:phen1}(b).}
	\label{tab:1}
\end{table*}
To substantiate the features of the different scenarios in terms of the velocity peak
values
$V_{p.1}$ and $V_{p.2}$, both are plotted in Figure \ref{fig:Fig3}(e) as a function of the distance $S_{E2}$,
which can be increased by choosing larger values $t_0$ of phase "0".
Furthermore, different radii $R_1$ of the first bubble are studied by varying the duration of $t_1$ between  1 ms $\leq t_1 \leq $5 ms. As can be seen, all three scenarios are examined. 

When $S_{E2}$ increases, $V_{p.1}$ monotonically falls from almost 60 mm/s at $S_{E2}\sim 175 \,\mu$m to reach a plateau of $V_{p.1} \sim 10...20 $ mm/s at large $S_{E2}$ values, at which the first bubble is by now barely affected by the second one.   
As $V_{p.1}$ results to a considerable extent from the rapid displacement of the electrolyte due to growth of the 2nd bubble, the existence of a plateau at $S_{E2}\approx S_{E2.6} \approx 375\, \mu$m demonstrates that a maximum distance must not be exceeded for an interaction between both bubbles.
$V_ {p.1}$ furthermore increases with the bubble radius $R_1$, and hence the buoyancy, for all the measurements performed.
On relating the plateau values to the corresponding terminal velocity 
at different $R_1$ we obtain: $(V_{p.1:min}/V_t)_{97\mu m}$=0.96;
$(V_{p.1:min}/V_t)_{80\mu m}$=1.15 and $(V_{p.1:min}/V_t)_{60\mu m}$=1.7. 
As seen in Table \ref{tab:1}, the terminal velocity is not quite achieved at $S_{E2} < 400$ $\mu$m. Hence, in all three cases ($R_{60, 80, 97}$), the velocity ratios demonstrate the acceleration of the first bubbles by the displaced electrolyte to velocities close to or higher than their terminal values.

The second velocity peak $V_{p2}$ behaves differently
and shows a parabola-like behavior at all bubble sizes $R_1$. 
When $S_{E2}$ increases, in the case of the smallest bubble $R_1=60\, \mu$m, 
$-V_{p2}$ rises to the maximum value of around 55 mm/s. This is followed by a gradual decline to zero, denoting the end of scenario II if $S_{E2}$ becomes too large. 
The local maximum is shifted to higher $S_{E2}$ values if the radius $R_1$ is increased.
The existence of a local maximum of $-V_{p2}$ indicates that the first bubbles, which are too far away, are only weakly influenced, while bubbles that are too close do not have the time to develop the local maximum of $V_{p2}$ before coalescing with the second bubble.

\begin{figure}
	\centering
	\includegraphics[width=0.8\textwidth]{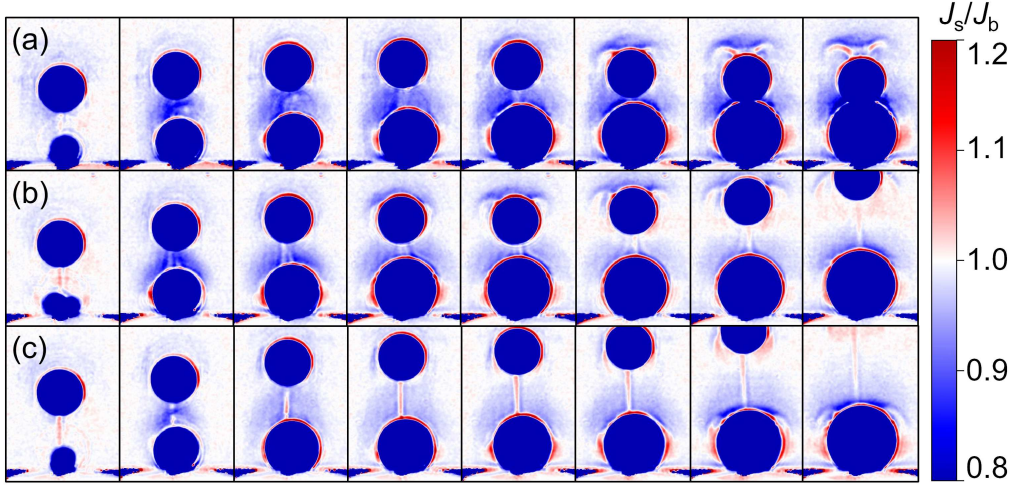}
	\caption{Schlieren images for different stages of scenarios I (a), II (b), and III (c), shortly after $E_2=-8$ V is switched on, taken at 
	$\Delta t= 3$ ms (a) and 4 ms (b-c). The
	intensity of the schlieren images $J_s$ is divided by $J_b$, the background schlieren image without bubbles.
		\label{fig:Fig4}}
\end{figure}

Using schlieren imaging, in Figure \ref{fig:Fig4} we analyze the vertical gradient of the refractive index, or temperature (cf. Section 2),  associated with the different scenarios I, II and III.
When the refractive index is translated into information on the temperature gradient, taking into account that the temperature is inversely proportional to the refractive index (\cite{haynes2016crc}), 
the red and blue colors in Fig. 4 denote an increase or decrease in the temperature in the vertical direction (from bottom to top).
Although the extent and instant of their appearance differs, Figure \ref{fig:Fig4} demonstrates the existence of blue regions on top of both bubbles for all three scenarios. This is reminiscent of the thermal boundary layer produced during bubble nucleation at the electrode 
on top of each bubble due to Joule heating. These boundary layers are advected 
during the rise of the first bubble and also during the growth of the second bubble. As a result, blue regions of decaying temperature are found near the top of the bubbles. 
It is also noticeable that the thermal schliere upstream of the second bubble rises faster due to the displaced electrolyte and the wake behind the first bubble.
As soon as this schliere reaches the bottom of the first bubble, the deceleration of the latter sets in. For sufficiently small distances $S_{E2}$, belonging to scenario I, a warmer region is established between the two bubbles. The resulting temperature gradient around the first bubble is
responsible for its downward acceleration towards the second bubble (cf. Section 4). As a result of this reverse motion, the thermal boundary layer on top of the first bubble develops a characteristic shape resembling a flying bird. 
If the bubble is further away from the electrode 
(scenario II), the schliere still touches the first bubble. However, the resulting temperature gradient is too weak to provoke an acceleration leading it to coalesce with the second bubble.
In scenario III, the blue zone of elevated temperature 
is too far from the first bubble,
hence there is no interaction at all.

\section{Discussion and conclusions}
\label{sec:disc}
The key phenomenon discovered in this work is the initially contactless interaction between
two electrogenerated bubbles, forcing a paradoxical reversal of the bubble motion in the opposite direction to buoyancy. The origin of this phenomenon needs to be sought in the forces, $F_e$ and $F_h$ (Section 1), acting on the bubble. 
Any influence from the electric force $F_e$ can be excluded for two reasons. (i) \cite{bashkatov2019oscillating} showed the non-linear dependency of $F_e$ and its strong decay for electrode distances larger than approx. $30 \,\mu$m. Thus, the electric force is unlikely to play any role at the much larger bubble-electrode distances of $\approx$ 300 $\mu$m found in this work. (ii) In another test experiment, a second bubble is produced in the same way as the first bubble at $E_2=-6$ V for 1 ms. Afterwards, $E_2$ is suddenly set to $E_2 = 0$ to force the detachment of the second bubble. Despite the vanishing potential $E_2$, $F_e=0$, the first bubble thus still follows scenario I in a nearly unchanged fashion.

This suggests that $F_h$, and in particular the Marangoni force element $F_M$, play a key role.
As revealed by the
schlieren images in Figure 4, the H$_2$ bubble motion reversal (scenarios I and II) sets in if the thermal boundary layer on top of the second bubble is able to touch the bottom of the first bubble.
 If this happens, a temperature gradient is built up along the surface of the first bubbles. As the surface tension $\gamma$
 decreases with increasing temperature, a 
$\gamma$ gradient 
 is established that pulls the adjacent electrolyte from the bubble foot towards the equator. As the  $\gamma$ gradient mimics the action of the arms of a swimmer, the bubble starts moving opposite to the buoyancy force. This is in analogy to the classical work by \cite{young1959motion} on thermocapillary migration with the exception that the temperature gradient is not externally imposed but generated in a self-organized way, including production by Joule heating and dissipation by thermal diffusion.
According to \cite{morick1993migration} and \cite{young1959motion}, the stationary velocity of the “creeping” thermocapillary migration of a bubble inside a temperature gradient $\partial T/\partial z$ is given by
\begin{equation}
 v = \frac{R^2 \rho g - (3/2)R(\frac{\partial \gamma}{\partial T} \frac{\partial T}{\partial z})}{3\eta}
\label{eq:migration_velocity}
\end{equation}
Taking the second peak velocity $V_{p2}\sim 30 $ mm/s as the characteristic velocity of
thermocapillary migration, the required $\partial T/\partial z$ according to Eq.~\ref{eq:migration_velocity}
is approx. 
90 K/cm. With the characteristic temperature rise $\Delta T \sim 10$ K at the microelectrodes (\cite{massing2019thermocapillary,hossain2020thermocapillary})
over a bubble radius $R \sim 100\:\mu$m (0.01 cm), a much larger $\partial T/\partial z \sim 10^3\:K/cm$ can easily be achieved.
This strongly supports the notion that (i) thermocapillary migration is at the origin of H$_2$ bubble motion reversal and (ii) the interaction of electrogenerated bubbles needs to be taken into account during water electrolysis.

\section{Acknowledgments}{This project is supported by the German Space Agency (DLR) with funds provided by the Federal Ministry of Economics and Technology (BMWi) due to an enactment of the German Bundestag under Grant No. DLR 50WM2058 (project MADAGAS II). We thank J. Boenke for his support with experiments during his student internship in 2021.}

\bibliographystyle{unsrtnat}
\bibliography{references}
\end{document}